%% file: QaryTR2arxiv.tex
\documentclass[10pt]{article}
\usepackage{color, palatino}

\usepackage[psamsfonts]{amssymb}
\usepackage{amsmath, amsthm, graphicx, enumerate, subfigure, color,bbm}
\usepackage{color}
\usepackage[top=3cm, bottom=3cm, left=3.5cm, right=3.5cm]{geometry}
\usepackage{algorithmic}
\usepackage{algorithm}
\usepackage{hyperref}
\usepackage[all]{hypcap}   
\usepackage{tikz}
\usepackage{tikz}
\usetikzlibrary{arrows,calc}
\usepackage{subfigure}
\usepackage{enumerate}
\input{macro}

\usepackage{lipsum}
\hypersetup{
    bookmarks=true,         
    unicode=false,          
    pdftoolbar=true,        
    pdfmenubar=true,        
    pdffitwindow=false,     
    pdfstartview={FitH},    
    pdftitle={My title},    
    pdfauthor={Author},     
    pdfsubject={Subject},   
    pdfcreator={Creator},   
    pdfproducer={Producer}, 
    pdfkeywords={keyword1} {key2} {key3}, 
    pdfborder={0 0 2 [3 3]},
    pdfnewwindow=true,      
    colorlinks=true,       
    linkcolor=red,          
    citecolor=green,        
    filecolor=magenta,      
    urlcolor=cyan           
}

\begin{document}

\title{q-ary Compressive Sensing}

\author{\normalsize{Youssef Mroueh$^{\top,\dagger}$, Lorenzo Rosasco$^{\dagger,\S}$}\\
\small \em $\top$  
CBCL, CSAIL,  Massachusetts Institute of Technology\\
\small \em $\dagger$ 
LCSL , Istituto Italiano di Tecnologia and IIT@MIT lab, Istituto Italiano di Tecnologia\\
\small \em $\S$ 
DIBRIS, Universita' degli Studi di Genova\\
\small \tt  \{ymroueh, lrosasco\}@mit.edu}

\maketitle


\maketitle

\begin{abstract}
We introduce $q$-ary compressive sensing,  an extension of $1$-bit compressive sensing. 
We propose a novel  sensing mechanism and a corresponding recovery procedure.
The recovery properties of the proposed approach are analyzed both theoretically and empirically. 
Results in $1$-bit compressive sensing are recovered as a special case. 
Our theoretical results suggest a tradeoff between the quantization parameter $q$, and the number of measurements $m$ in the control of the error of the resulting recovery algorithm, as well its robustness to noise. 
\end{abstract}

\section{Introduction}
%
%
Reconstructing signals from discrete measurements is a classic problem in signal processing.
Properties of the signal inform the way reconstruction can be achieved from a minimal set of measurements.
The classical Shannon sampling result ensures that band limited signals can be reconstructed by a linear procedure, 
as long as a number of  linear measurements, at least twice the maximum frequency, is available. Modern data analysis typically requires recovering  high dimensional signals from few inaccurate measurements. Indeed, the development of   Compressed Sensing (CS) and Sparse Approximation \cite{survey}
shows that   this is  possible  
for signals with further structure.
For example,   $d$-dimensional, $s$-sparse signals\footnote{A $d$-dimensional signal, that is a vector in ${\mathbb R} ^d$, is  $s$-sparse if only $s$ of its components are different from zero.} 
can be reconstructed with high probability through convex programming, given $m\sim s\log(d/s)$ random linear 
measurements. 

Non linear measurements have been recently considered in the context of $1$-bit compressive sensing ({\small \texttt{http://dsp.rice.edu/1bitCS/}}). 
Here, binary (one-bit) measurements are obtained by applying, for example, the ``sign'' function\footnote{More generally,  any function  $\theta:\R\to [-1,1]$, such that $\mathbb{E}(g\theta(g)>0)$ can be used. } 
to linear measurements. More precisely, given 
$x\in \R^{d}$,  a measurement vector is given by  $y=(y_1, \dots, y_m)$, where $y_i= sign(\scal{w_i}{x})$ with  $w_i\sim {\mathcal N} (0, I_{d})$ independent Gaussian random  vectors, for $i=1,\dots, m$. It is possible to prove \cite{Vershynin}  that, for a signal $x \in K\cap  \mathbb B ^{d}$ ($\mathbb{B}^{d}$ is the unit ball in $\R^d$),
 the solution $\hat x_m$ to the problem
\begin{equation}\label{ERM}
\max_{x\in K}\sum_{i=1}^m y_i\scal{w_i}{x},
\end{equation}
satisfies $\nor{\hat x_m - x}^2\le  \frac{\delta}{\sqrt{\frac{2}{\pi}}}$, with probability $1-8\exp{(-c\delta^2m)}$, $\delta>0$, as long as $m\ge  C\delta^{-2} \omega(K)^2$ \cite{Vershynin}. Here, $C$ denotes a universal constant and   $\omega(K)=\E \sup_{x\in K-K} \scal{w}{x}$  the Gaussian mean width $K$, which  can be interpreted as a complexity measure. If $K$ is a convex set, problem \eqref{ERM} can be solved efficiently. 

In this paper, borrowing ideas from signal classification studied in machine learning, we discuss   
a novel sensing strategy, based on  \emph{$q$-ary non linear measurements}, 
and  a corresponding recovery procedure. 

\section{q-ary Compressive Sensing}
In this section we first describe the sensing and recovery procedure (Section~\ref{sensrec}), then describe the results in the noiseless (Section~\ref{nonoise}) and noisy setting (Section~\ref{noise}), and finally, we sketch the main ideas of the proof (Section~\ref{proofsk}).

\subsection{Sensing and Recovery}\label{sensrec}
%
 The sensing procedure we consider is given  by  a map $C$ from $K\cap {\mathbb B}^{d}$  to the $q$-ary , $m$ Hamming cube $\{0,\dots, q-1\}^{m}$, where $K\subset \R^d$ 
To  define  $C$ we need the following definitions. 
\begin{definition}[Simplex Coding \cite{Mclass}]\label{scode}
The simplex coding map is $S:\{0,\dots, q-1\} \to \R^{q-1}$,  
$\quad S(j)=\a_j,$ where\\ 
1) $\norT{\a_j}^2=1$, \\
2)$\scalT{\a_j}{\a_{i}}=-\frac{1}{q-1},$ for $ i\neq j$, \\
3) $\sum_{j=0}^{q-1}\a_j =0$,\\
for all $i,j=0, \dots, q-1$.

\end{definition}
\begin{definition}[$q$-ary Quantized Measurements] 
Let 
 $W \in \mathbb{R}^{q-1,d}$ be a  Gaussian random matrix, i.e. $W_{ij} \sim \mathcal{N}(0,1)$ for all $i,j$.
Then, 
$Q: K\cap {\mathbb B}^{d} \to \{0,\dots,q-1\}$,$$ Q(x)=Q_{W}(x)=\argmax_{j=0\dots q-1} \scalT{s_j}{Wx},$$
is called a  $q$-ary quantized measurement.
\label{def:quant}
\end{definition}
\noindent Then, we can define the $q$-ary sensing strategy induced by  non linear quantized measurements. \begin{definition}[$q$-ary  Sensing]\label{qarySense}
Let $W_1,\dots, W_m$, be  independent Gaussian random matrices in $\mathbb{R}^{q-1,d}$ and 
 $Q_{W_i}(x),i=1,\dots, m$  as in Def. \ref{def:quant}.  
The $q$-ary  sensing is $C: K\cap {\mathbb B}^{d} \to\{0,\dots,q-1\}^m$, 
$$C(x)=(Q_{W_1}(x),\dots Q_{W_m}(x)),$$
 $\forall x \in K\cap{\mathbb B} ^{d}$.
\end{definition}
\noindent Before describing the recovery strategy we consider, we add  two  remarks.
\begin{remark}[Connection to $1$-bit CS]
If $q=2$, $W$ reduces to a  Gaussian random  vector, and  $2 Q(x)-1=sign(Wx)$,
so that the $q$-ary quantized measurements  become equivalent to those 
considered in in $1$-bit CS.
\end{remark}
\begin{remark}[Sensing and Embeddings]
It can be shown  that  $C$ defines an $\epsilon$-isometric embedding of $(K,\nor{\cdot})$, into $(\mathcal{F},d_{H})$ -- up-to a bias term.
Here $d_{H}$ is the (normalized) Hamming distance,  $d_{H}(u,v)=\frac{1}{m}\sum_{i=1}^m \mathbbm{1}_{u_i\neq v_i}$,$u, v \in \mathcal{F}$. This analysis  is deferred to the  long version of this paper.
\end{remark}

In this paper, we are interested in provably (and  efficiently) recovering  a signal $x$ 
from  its $q$-ary measurements $y=(y_1, \dots,y_m)=C(x)$.
Following \cite{Vershynin}, we consider the recovery strategy $D: \{0,\dots,q-1\}^m\to K\cap{\mathbb B} ^{d}$  defined by, 
\begin{equation}
D(y)=\argmax _{u \in K \cap \mathbb{B}^{d}} \frac{1}{m} \sum_{i=1}^ m \scalT{\a_{y_i}}{W_i u}.
\label{eq:decoder}
\end{equation}
The above problem is convex as soon as $K$ is convex and can be solved efficiently, see Section~\ref{sec:recov}.
In the next section, we prove it has good recovery guarantees both in noiseless and noisy settings.
We first add a  remark. 

%

 \begin{remark}[Connection to Classification]\label{SVMDecode}
An inspiration for  considering the $q$-ary CS stems from an analogy between $1$-bit compressed sensing and  binary classification in machine learning. In this view, Definition~\eqref{qarySense} is  related to the approach  proposed  for multi category classification in \cite{Mclass}.  Following these ideas, we can extend the recovery strategy~\eqref{eq:decoder} by considering 
\begin{equation}
 D_{V}(y)=\argmin_{u \in K \cap \mathbb{B}^{d}} \frac{1}{m}\sum_{i=1}^m V(-\scalT{\a_{y_i}}{W_iu}),
 \end{equation}
where $V$ is a convex,  Lipchitz,   non decreasing  loss function $V:\mathbb{R}\to \mathbb{R}^+$. 
Problem~\eqref{eq:decoder} corresponds   to the choice $V(x)=x$. Other possible choices include $V(x)=\max(1+x,0)$, $V(x)=\log(1+e^x)$, and $V(x)=e^x$.
All these loss functions can be seen as  convex relaxations of the  $0$-$1$ loss function, 
defined as $V(x)=0$ if $x\leq 0$, and $1$ otherwise. The latter defines the misclassification risk,  which corresponds to Hamming distance in CS,  which is the natural measure of performance while learning classification rules. 
%
 \end{remark}

\subsection{Recovery guarantees: Noiseless Case }\label{nonoise}
The following theorem describes the recovery guarantees for the proposed procedure, when applied on a signal $x$  in  a set $K$ of Gaussian mean width $w(K)$. We first consider a noiseless scenario.

\begin{theorem}
 Let $\delta >0$, and 
$m\geq C\delta^{-2}w(K)^2.$
Then with probability at least $1-8\exp(-c\delta^2 m)$, the solution $\hat{x}_m=D(y)$ of problem \eqref{eq:decoder} satisfies,\\
 \begin{equation}\label{theo2}
 \nor{\hat{x}_m-x}^2\leq \frac{\delta}{\sqrt{\log(q)}}.
 \end{equation}
\label{theo:theorem2}
\end{theorem} 
\noindent A proof sketch of the above result   is given in Section \ref{proofsk}, 
while the complete proof is deferred to the long version of the paper. 
Here, we add four comments. 
First, we note that the above result implies the  error bound, 
 \begin{equation}
||\hat{x}_m-x||^2\leq C(\frac{w(K)}{\sqrt{\log(q)m}}+\delta),
\label{eq:tradeoff}
\end{equation}
with probability at least, $1-4\exp(-2\delta^2),\delta>0$.

Second, Inequalities~\eqref{theo2},~\eqref{eq:tradeoff} can be compared 
to results in $1$-bit CS.  For the same number of measurements, 
$m\geq C\delta^{-2}w(K)^2$, the error for $q$-ary CS is  $\frac{\delta}{\sqrt{\log(q)}}$, in contrast with $\frac{\delta}{\sqrt{\frac{2}{\pi}}}$ in the $1$-bit CS \cite{Vershynin}, at the expense of 
a more demanding  sensing procedure. Also note that, for $q=2$, we  recover the result in $1$-bit CS as a special case.
Third, we see that for a given accuracy our results highlights a trade-off between 
the number of $q$-ary measurements $m$ and the quantization parameter $q$.
To achieve an error $\epsilon$, with a memory budget of $\ell$ bits, one can choose  $m$ and $q$ so that  $\epsilon= O(\frac{1}{\sqrt{m\log(q)}})$, and $m\log_{2}(q)= \ell$ (see also  section \ref{sec:tradeoff}).
Finally, in the following we will be interested in $K$ being the set of $s$-sparse signals.
Following again \cite{Vershynin}, it is interesting to consider in Problem~\eqref{eq:decoder} the relaxation 
$$K_1=\{u\in \mathbb{R}^d: \nor{u}_{1}\leq \sqrt{s}, \nor{u}_2\leq1\}.$$ 
With this choices, it it possible to prove that $w(K_1)\leq C\sqrt{s\log(\frac{2d}{s})}$, and  that 
for $m\geq C\delta^{-2} s\log(\frac{2d}{s})$, the solution of the convex program \eqref{eq:decoder} on $K_1$ satisfies, $\nor{\hat{x}_m-x}^2\leq \frac{\delta}{\sqrt{\log(q)}}$.
We end noting that other choices of $K$ are possible, for example  in \cite{Groupsparse} the set of group sparse signals (and their Gaussian width) are studied.

 \subsection{Recovery Guarantees: Noisy Case}\label{noise}
Next we discuss
the $q$-ary approach in two noisy settings,
related to those considered  in  \cite{Vershynin}.

\noindent {\bf  Noise before quantization.} 
For $i=1,\dots, m$, let 
\begin{equation}
\label{nbq}
y_i = \arg\max_{j=0\dots q-1}\{\scalT{s_j}{W_ix}+g_j\},
\end{equation} 
with  $g_j$ independent Gaussian realization of variance  $\sigma^2$.
 In this case, it is possible to prove that, for $m \geq C\delta^{-2}w(K)^2$, 
 $$\nor{\hat{x}_m-x}^2\leq \frac{\delta\sqrt{1+\sigma^2}}{\sqrt{\log(q)}},$$
 with probability at least $1-8\exp{(-c\delta^2m)}$.
The  quantization level $q$ can be chosen  to adjust to  the noise level $\sigma$   for a more robust recovery of $x$. This result can be viewed in 
the  perspective of the {\em bit-depth versus measurement-rates} perspective studied in \cite{Baraniuk}. Here it is  shown  that $1$-bit CS outperforms conventional scalar quantization. In this view, $q-$ary CS provides  a new way to adjust the quantization parameter  to the noise level. \\
\noindent {\bf Inexact maximum.} 
For $i=1,\dots, m$, let  $y_i=Q_{W_i}(x)$, with probability $p$, and $y_i=r$ with probability $1-p$, with $r$  drawn uniformly  at random from $\{0,\dots, q-1\}$. 
In this case, it is possible to prove that, for $m \geq C\delta^{-2}w(K)^2$, 
$$
\nor{\hat{x}_m-x}^2\leq \frac{\delta}{\sqrt{\log(q)}(2p-1)}.
$$
with probability at least $1-8\exp{(-c\delta^2m)}$. 
The signal $x$ can be recovered even if {\em half} of the $q$-ary bits are flipped.
\subsection{Elements of the proofs}\label{proofsk}
We sketch the main steps in proving our results. 
The proof of Theorem \ref{theo:theorem2} is based on: 1) deriving a bound in expectation, 
and 2)  deriving a concentration result. The proof of the  last step 
uses Gaussian concentration inequality  extending  the proof strategy in \cite{Vershynin}. 
Step 1) gives the bound
$$
\mathbb{E} \left(||\hat{x}_m-x||^2\right)\leq  \frac{ w(K)}{C\sqrt{\log(q)m}},
$$
the proof of which is based on the following proposition.  
\begin{proposition}Let $\mathcal{E}_{x}(u)=\mathbb{E}_{W}(\scalT{\a_{\gamma}}{Wu})$, where $\gamma=Q_{W}(x)$. Then, $\forall u \in \mathbb{B}^{d}$, we have,
$$
\frac{1}{2}\nor{u-x}^2\leq \frac{1}{\lambda(q)} \left(\mathcal{E}_x(x)- \mathcal{E}_x(u)\right), 
$$
where $\lambda(q)=\mathbb{E}_{\bar{\gamma},g}(\scalT{\a_{\bar{\gamma}}}{g})$, and $g \sim \mathcal{N}(0,I_{q-1})$, and $\bar{\gamma}=\argmax_{j=0\dots q-1} \scalT{s_j}{g}$. 
\label{lem:lemma1}
\end{proposition}
\noindent Using results in 
empirical process theory it possible  to show that   
$$
|\mathcal{E}_x(x)- \mathcal{E}_x(\hat{x}_m)|
 \le C \frac{w(K)}{\sqrt{m}}.
$$
The  bound on the expected recovery follows combining the above inequality and Proposition~\ref{lem:lemma1} with the inequality, 
$$\lambda(q)\geq C\sqrt{\log(q)},$$
which is proved using 
 Slepian inequality and Sudakov minoration. 

The results in the noisy settings follow from suitable estimates of  $\lambda(q)$.
Indeed, for the {\em noise before quantization} case it can be  proved that  $\lambda(q)\geq C\sqrt{\frac{\log(q)}{1+\sigma^2}}$. For the {\em inexact maximum} case
one has 
\begin{eqnarray*}
\lambda(q)&&=\mathbb{E}_{y,g}(\scalT{s_y}{g})=\\
&&p \mathbb{E}(\max_{j=1\dots q}\scalT{s_j}{g})+(1-p)\mathbb{E}(\scalT{s_{r}}{g})\geq \\
&&Cp\sqrt{\log(q)}+(1-p) \mathbb{E}(\min_{j=1\dots q} \scalT{s_j}{g})\geq \\
&&(2p-1)C\sqrt{\log(q)}.
\end{eqnarray*}

\section{Experimental Validation}
In this section, we describe some numerical simulations in sparse recovery, Section~\ref{sec:recov}, and  preliminary experiments in an image recovery problem, Section~\ref{image}.
\subsection{An Algorithm for Sparse recovery}\label{sec:recov}
In our experiments, we considered the following variation of  problem \eqref{eq:decoder}, 
Let $\xi_{i}=\a_{y_i}^{\top}W_i \in \mathbb{R}^d, i=1\dots m$.
\begin{equation}
\max_{u,\nor{u}_{2}\leq1} \frac{1}{m}\sum_{i=1}^m \scalT{\xi_{i}}{u}-\eta \nor{u}_{1},
\label{eq:sparserecov}
\end{equation}
where $\eta>0$.
The above problem can  be solved efficiently using Proximal Method \cite{convex},  a 
solution can be computed via the iteration,
\begin{eqnarray*}
u_{t+1}&=& u_{t} +  \frac{\nu_{t}}{m}\sum_{i=1}^m \xi_{i},\\
u_{t+1}&=&Prox_{\eta} (u_{t+1}),\\
u_{t+1}&=&u_{t+1}\min (\frac{1}{\nor{u_{t+1}}_{2}},1).
\end{eqnarray*}
Where $\nu_{t}$ is the gradient step size, and $Prox_{\eta }$ acts component 
wise as $\max(1-\frac{\eta}{|u_i|},0)u_i$.
The iteration is initialized randomly to a unit vector.

\begin{remark}
The computational complexity of the sensing process depends on both $m$ and $q$.
Whereas, the computational complexity of the recovery algorithm,
once computed $\xi_{i}$,  is independent to the choice of $q$, and depends only on $m$
and is the same as in $1$-bit CS.
\end{remark}

\subsection{Sparse Recovery}\label{sec:tradeoff}
\begin{figure}[H]
\subfigure[Error $\nor{x-\hat{x}}^2$ versus q, for  $m=70, d=100$.]{
\includegraphics[scale=0.3]{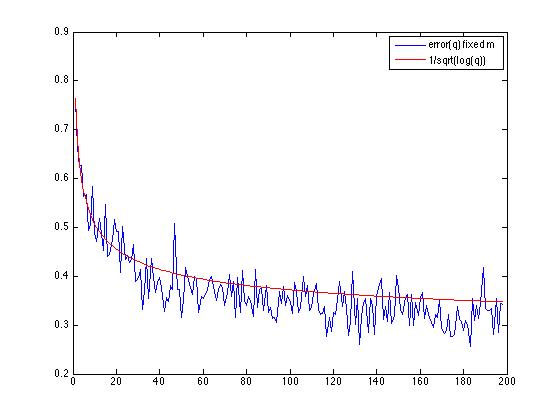}
\label{fig:subfig2}
}
\hspace{.7cm}
\subfigure[Error $\nor{x-\hat{x}}^2$ versus $m$, for $q=3, d=100$.]{
\includegraphics[scale=0.3]{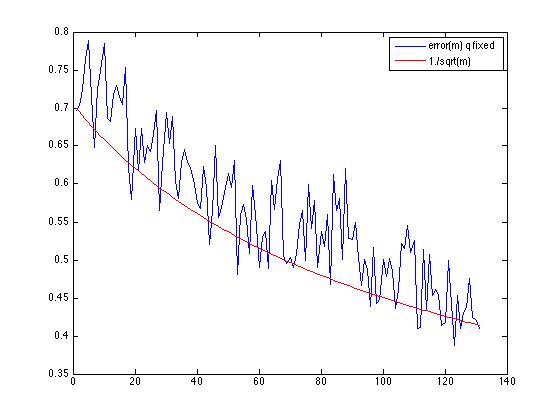}
\label{fig:subfig3}
}  \\
\subfigure[Theoretical bound for  $\nor{x-\hat{x}}^2$ versus $m$ and $q$.]{
\includegraphics[scale=0.3]{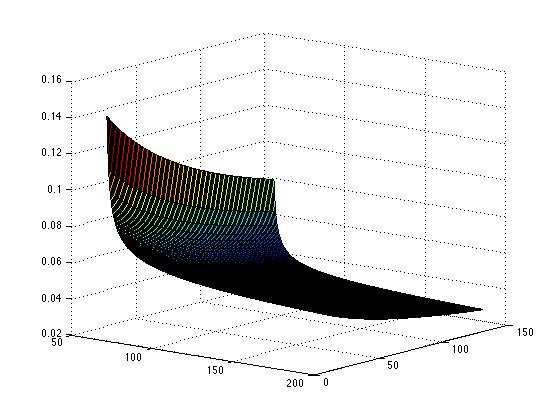}
\label{fig:subfig4}
}
\hspace{.7cm}
\subfigure[Error $\nor{x-\hat{x}}^2$ versus $m$ and $q$.]{
\includegraphics[scale=0.3]{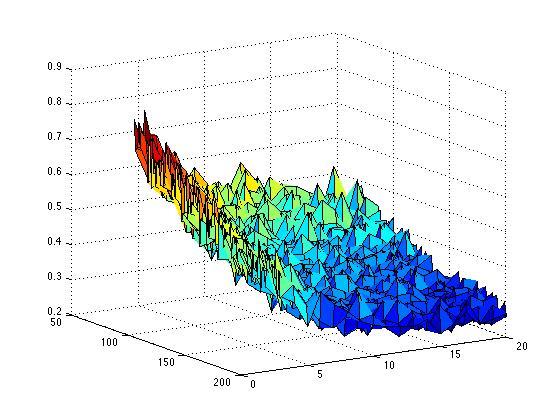}
\label{fig:subfig5}
}
\label{fig:subfigureExample}
\caption{q-ary Compressive Sensing: Quantization/Number of measurements tradeoff.}
\end{figure}

We tested our approach for  recovering  a signal from from its $q$-ary measurements. 
We considered sparse signals of dimension $d$ generated via a Gauss-Bernoulli model.
In Figure \ref{fig:subfig2}, we see that the  reconstruction error of $\hat{x}_m$ (in blue), for varying   $q$ and $m$ fixed, follows the theoretical bound $\frac{1}{\sqrt{\log(q)}}$ (in red). In Figure \ref{fig:subfig3}, we see that the reconstruction error $\hat{x}_m$ (in blue),  for varying   $m$ and $q$ fixed, follows  the theoretical bound $\frac{1}{\sqrt{m}}$ (in red). Figures \ref{fig:subfig4}, and \ref{fig:subfig5} highlight the tradeoff between the number of measurements and the quantization parameter. For a precision $\epsilon$, and a memory budget  $2^B$, one can choose an operating point $(m,q)$, according to the theoretical  bound $\frac{1}{\sqrt{m\log(q)}}$.

\subsection{Image Reconstruction}\label{image}
\begin{figure}[H]
\begin{center}
\includegraphics[scale=0.45]{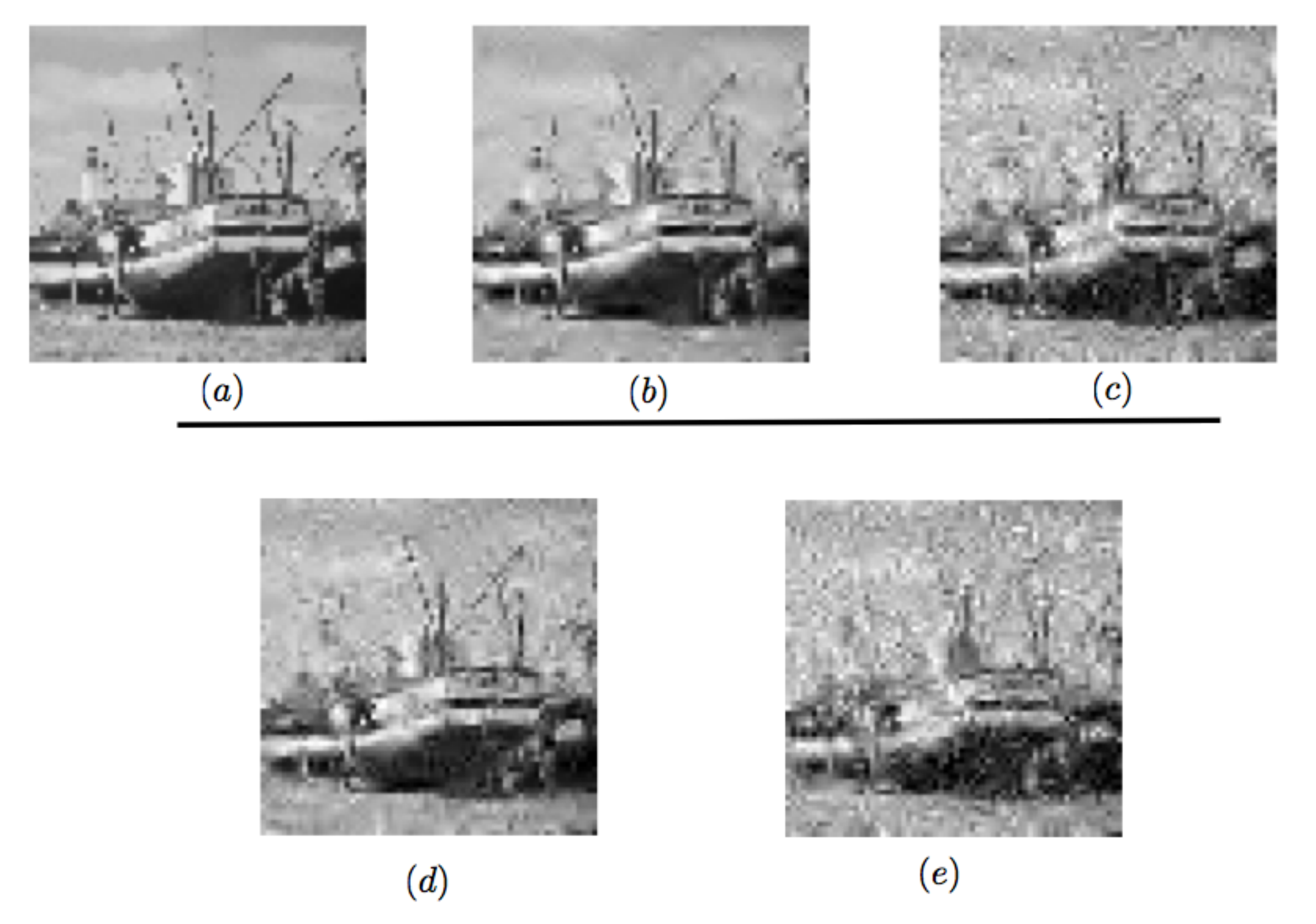}
\end{center}
\caption{Image recovery with $q$-ary CS. (a) Original image. 
(b)  Reconstruction with no-noise: $q=2^5$,   $\text{ SNR}=20.2$ dB.
(c)  Reconstruction with no-noise:   $q=2$, $\text{ SNR }=16.2$ dB.
(d) Reconstruction with noise: $q=2^5$, $\text{ SNR }=18.3$ dB.
(e) Reconstruction with noise: $q=2$, $\text{ SNR }=15$ dB.}
%
%
\label{fig:Image}
\end{figure}

Then, we  considered the problem of recovering an image  from $q$-ary measurements. We used   the standard $8-$bit grayscale boat  image of size $64\times 64$ pixels shown in Figure~\ref{fig:Image}\textcolor{red}{(a)}.
We extracted the wavelet coefficients and performed thresholding  to get a sparse signal. We normalized the resulting vector of wavelets coefficients of dimension $d=3840$ to obtain a unit vector. Then, we performed sensing and recovery with 
 $q=2^5$ ($5$-bit compressive sensing ) and $q=2$ ($1$-bit compressive sensing)
 for the same $m=2048<d$. We compared the SNR performances of the corresponding reconstructed images in a  noiseless setting (Figures~\ref{fig:Image}\textcolor{red}{(b)}-\textcolor{red}{(c)}), 
 and a noisy setting, considering the noise before quantization model ~\eqref{nbq},  with  $\sigma=0.8$
 (Figures~\ref{fig:Image}\textcolor{red}{(d)}-\textcolor{red}{(e)}. )
The results confirm our theoretical:  higher quantization improves the SNR, as well as  the robustness to noise of $q$-ary compressive sensing.

\end{document}

%% file: macro.tex
\makeatletter
\def\@makechapterhead#1{
\vspace*{80\p@}
{\parindent \z@ \raggedright \normalfont
\ifnum \c@secnumdepth >\m@ne
\if@mainmatter
\huge\bfseries \@chapapp\space \thechapter
\par\nobreak
\vskip 20\p@
\fi
\fi
\interlinepenalty\@M
\Huge \bfseries #1\par\nobreak
\vskip 40\p@
}}
\makeatother

\def\argmax{\operatornamewithlimits{arg\,max}}
\def\argmin{\operatornamewithlimits{arg\,min}}

\DeclareMathOperator{\E}{\mathbb{E}}

\renewcommand{\a}{s}

\newcommand{\iii}{\begin{enumerate}}
\newcommand{\fff}{\end{enumerate}}
\newcommand{\iiii}{\begin{itemize}}
\newcommand{\ffff}{\end{itemize}}
\newcommand{\mfi}{\begin{eqnarray*}}
\newcommand{\mff}{\end{eqnarray*}}
\newcommand{\mfni}{\begin{eqnarray}}
\newcommand{\mfnf}{\end{eqnarray}}

\newtheorem{theorem}{Theorem}

\newtheorem{definition}{Definition}

\newtheorem{proposition}{Proposition}
\newtheorem{remark}{Remark}

\providecommand{\nor}[1]{\left\lVert {#1} \right\rVert}

\providecommand{\scal}[2]{\left\langle{#1},{#2}\right\rangle}

\providecommand{\scalT}[2]{\left\langle{#1},{#2}\right\rangle}
\providecommand{\norT}[1]{\left\lVert {#1} \right\rVert}

\newcommand{\R}{\mathbb R}